\documentclass[prl,aps,twocolumn,showpacs]{revtex4}
\usepackage{txfonts}
\usepackage{bbm}
\usepackage{amssymb}
\usepackage{graphicx}
\usepackage{subeqnarray}

\usepackage{color}

\begin{document}
\title{Functionalized Bismuth Films: Giant Gap Quantum Spin Hall and Valley-Polarized Quantum Anomalous Hall States}
\author{Chengwang Niu}
\email{c.niu@fz-juelich.de}
\author{Gustav Bihlmayer}
\author{Hongbin Zhang} 
\author{Daniel Wortmann} 
\author{Stefan Bl\"{u}gel}
\author{Yuriy Mokrousov}
\affiliation{ Peter Gr\"{u}nberg Institut and Institute for Advanced Simulation, Forschungszentrum J\"{u}lich and JARA, 52425 J\"{u}lich, Germany}          

\begin{abstract}
The search for new large band gap quantum spin Hall (QSH) and quantum anomalous Hall (QAH) insulators is critical for 
their realistic applications at room temperature. Here we predict, based on first principles calculations, that the band gap of QSH and QAH states can be as large as 1.01 eV and 0.35 eV in an H-decorated Bi(111) film. The origin 
of this giant band gap lies both in the large spin-orbit interaction of Bi and the H-mediated exceptional electronic and structural properties. Moreover, we find that the QAH state also possesses the properties of quantum valley Hall state, thus intrinsically realising the so-called valley-polarized QAH effect. We further investigate the realization of large gap QSH and QAH states in an H-decorated Bi(\={1}10) film and X-decorated (X=F, Cl, Br, and I) Bi(111) films.  
\end{abstract}

\maketitle
\date{\today}

Since their discovery~\cite{kane,bernevig}, there is growing interest in topological insulators (TIs), which host 
conducting surface states inside the bulk insulating gap. The gapless surface states are topologically protected by 
time reversal symmetry (TRS) and robust to nonmagnetic perturbations~\cite{moore,hasan,qi}. The first theoretically 
predicted~\cite{bernevig2} and experimentally observed~\cite{konig} TI is a HgTe/CdTe quantum well structure that is 
a two-dimensional (2D) TI, also known as quantum spin Hall (QSH) insulator. In a QSH insulator, pairs of 
dissipationless edge channels with opposite spins exist, leading to extraordinary properties and possible applications 
in low dissipation electronic devices. On the other hand the realization of the Quantum Anomalous Hall (QAH) effect, which 
was first suggested to occur in a honeycomb lattice model~\cite{haldane}, has been achieved recently in Cr-doped 
topological insulators (Bi,Sb)$_2$Te$_3$~\cite{chang} via suppressing one of the spin channels~\cite{liu,yu}, but requires extremely low temperatures (30 mK). For obtaining the room temperature QSH- and QAH-based electronic devices, searching for novel materials with large band gaps as well as stable atomic and magnetic structures has been a fairly important topic in the field. In spite of extensive efforts so far~\cite{liucc,Murakami,liuz,ZhangBi1,ZhangBi2,xu,wu,Weng,liuc,hzhang,zhanghb,qiao,wang}, most of known
systems with desired topological properties have a small band gap, which greatly obstructs their potential room temperature applications.

Valley polarization, as a new degree of freedom in honeycomb lattices in addition to the intrinsic charge and spin, has received considerable attention in recent years~\cite{dxiao,Mak}. The valley Hall conductivity can be non-zero when the inversion symmetry is broken, realizing the quantum-valley Hall (QVH) effect characterized by so-called valley Chern number~\cite{dxiao}. 
Quite recently, a new quantum state, valley-polarised QAH state that exhibits the electronic properties of both QVH state and QAH state has been predicted in silicene through tuning the extrinsic spin-orbit coupling (SOC) with broken TRS~\cite{pan}. It provides a new way to design the dissipationless valleytronics. However, the presence of both, inversion symmetry and TRS, as well as the small SOC in pristine silicene makes experimental studies and possible applications difficult. 

\begin{figure}[b!]
\includegraphics{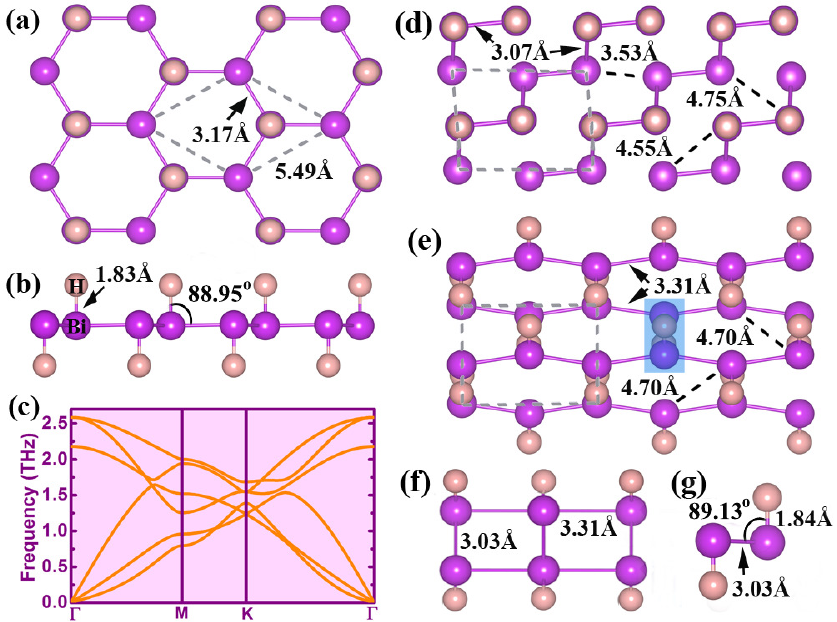}
\caption{(color online)
 Top (a) and side (b) view of optimized structures for H-Bi(111). The corresponding phonon band structure is
 shown in (c). Top view of the unrelaxed and relaxed crystal structure of H-Bi(\={1}10) is shown in (d) and (e),
 respectively. (f) Side view of the  relaxed crystal structure of H-Bi(\={1}10). (g) Zoomed-in view of the
 highlighted areas in (e). The unit cells are indicated by gray, dashed lines. The numbers are interatomic
 distances ($d_{\rm Bi-H}$ or $d_{\rm Bi-Bi}$) and angles ($\theta_{\rm H-Bi-Bi}$) between Bi (H) atoms.}
\label{structure}
\end{figure}

Generally, materials with strong SOC and simultaneously broken inversion and time-reversal
symmetries, which
exhibit non-trivial topological phases, are in high demand. As the heaviest atom with effectively stable isotope and strong SOC~\cite{Marcillac}, bismuth is an 
important ingredient for both 2D and 3D TIs, such as Bi(111) bilayer~\cite{Murakami,liuz}, Bi$_{1-x}$Sb$_x$~\cite{Fu,Hsieh}, Bi chalcogenides~\cite{Zhang,Xia}, and TlBiSe$_2$~\cite{Lin,Sato,Kuroda}. Bi(111) bilayer has drawn 
much attention due to a relatively large band gap of the 2D system ($\sim$0.2 eV~\cite{Murakami}) with the edge states observed experimentally~\cite{Drozdov}. Rather recently, Bi(111) bilayers
were grown on different substrates~\cite{Hirahara,zfwang,Kim}. On weakly interacting substrates ultrathin, (111) oriented films are unstable with respect to transformation into another
allotrope of Bi~\cite{Nagao}. It grows in the black-phosphorous (A17) structure, that resembles (110) layers of the bulk
Bi (A7) structure~\cite{Koroteev} and turns out to be topologically trivial~\cite{Wada}. Recently, atomic hydrogen 
chemisorbed on 2D TIs, for example graphene~\cite{sofo,elias,zhou} and stanene~\cite{xu}, has 
proved to be an effective way to modulate their properties. However, the band gap of graphene is quite small, and stanene is not a 2D TI in its equilibrium structure.

In this work, based on first-principles calculations, we show that both the electronic and topological properties of 
ultrathin Bi films can be drastically modified when decorated by H. A H-decorated Bi(111) film (H-Bi(111)) 
exhibits a topological energy gap of 1.01 eV, that is much larger than those in known TIs. Besides the strong SOC~\cite{song}, the lattice parameters also play critical role in determining the giant band gap of the 2D system. For the case of Bi(\={1}10) 
film (H-Bi(\={1}10)), H-decorating induces the realization of 2D TIs phase with a direct gap of 0.34 eV. Based 
on Chern number calculations, we further demonstrate that the valley-polarized QAH effect can be realized in Bi thin films hydrogenated only on one side of the films. 

\begin{figure}[t!]
\includegraphics{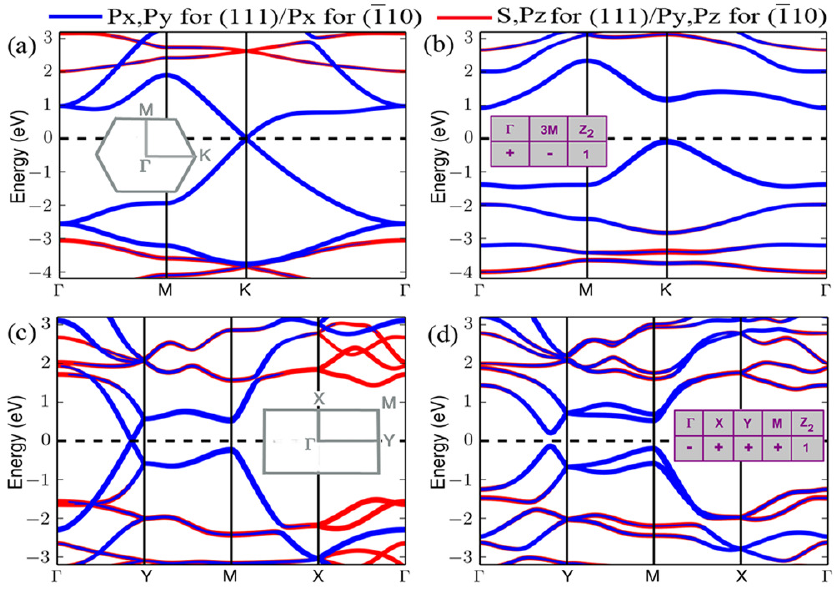}
\caption{(color online)
 Orbitally-resolved band structures for H-Bi(111) (a, b) and H-Bi(\={1}10) (c, d) without  (a, c) and with (b, d) SOC, 
 weighted with the s, p$_x$, p$_y$, and p$_z$ characters. Dark (blue) colors mark states that contribute to the
 fundamental band gap. The Fermi level is indicated by the dashed line. Insets in panel (a) and (c) show 
 the 2D Brillouin zone, and those in (b) and (d) show products of the parities of 
 all occupied bands at the time reversal invariant momenta and the $\mathbb{Z}_2$ number.}
\label{band}
\end{figure}

Bulk bismuth shows two different bond lengths: 3.07{\AA} within (111) bilayers and 3.53{\AA} between 
them. Projected onto the (111) plane, Bi bilayer forms a hexagonal lattice. In the Bi(\={1}10) 
layers, both two different bonds exist within a slightly buckled pseudosquare lattice. The crystal structures 
of H-Bi(111) and H-Bi(\={1}10) are plotted in Fig.~\ref{structure}. In contrast to decorated 
Sn~\cite{xu} , the H-Bi(111) has a quasi-planar geometry with the angle $\theta_{\rm H-Bi-Bi}$ 
being slightly smaller than $90^{\circ}$. Therefore, the hexagonal lattice parameter increases from 4.54{\AA} to 5.49{\AA},
while $d_{\rm Bi-Bi}$ is only expanded to 3.17{\AA}. H-Bi(\={1}10) forms a structure with an $AB$-stacking of the pseudo-square layers with $d_{\rm Bi-Bi}$ of 3.03{\AA} connecting the atoms within these layers and 3.31{\AA} between the layers. In addition, it is interesting to note that Bi and H atoms within the same layer have a configuration quite similar to that of H-Bi(111). The H-decoration changes the structure significantly, and the inversion symmetry is obtained as shown in Figs.~\ref{structure}(d) and \ref{structure}(e). We further confirm the stability of H-Bi(111) by phonon calculations. The real phonon frequencies at all momenta shown in Fig.~\ref{structure}(c) confirm that the structures are stable.

The calculated band structures for H-Bi(111) and H-Bi(\={1}10) are plotted in Fig.~\ref{band}. In the case without SOC, as 
shown in Figs.~\ref{band}(a) and (c), the systems are gapless and show semimetallic character with one band crossing 
exactly at the $K$ point for the H-Bi(111) case and slightly away from the $Y$ point for the case of H-Bi(\={1}10). This is different from other Bi-based TIs, such as the Bi(111) bilayer~\cite{Murakami} or the Bi$_2$Se$_3$~\cite{Zhang}, but quite similar to graphene~\cite{kane}. Taking SOC into account, a band gap opens (Figs.~\ref{band}(b) and (d)). Different from both the Bi-based TIs and graphene, the Dirac-related bands have mainly contributions from p$_x$ and p$_y$ orbitals while the p$_z$ orbital is removed away from the Fermi level by H, resulting in the large band gap. Similar mechanism was reported recently for Bi/Si system~\cite{zhoum}. To identify the band topology, the $\mathbb{Z}_2$ invariant is investigated by evaluating the wave function parities at four time reversal invariant momentum (TRIM) points~\cite{Fu}, i.e.\ the $\Gamma$ and three $M$ points for H-Bi(111) and $\Gamma$, $X$, $Y$, and $M$ for H-Bi(\={1}10) (insets of Figs.~\ref{band}(a) and (c)). The product of the parities of the occupied bands at TRIM $k$, $\delta_k$, is given in the insets of  Figs.~\ref{band}(b) and (d), together with the $\mathbb{Z}_2$ number ($v$) determined by $(-1)^{v} = \prod_k \delta_k$. We verify that both the H-Bi(111) and H-Bi(\={1}10) are QSH insulators (not for H-Bi(110) and H-Bi(A17)). The QSH state is further explicitly confirmed by the emergence of the gappless edge states in thin nanoribbons of the bilayers~\cite{sup}. 

\begin{figure}[htbp]
\centering
\includegraphics{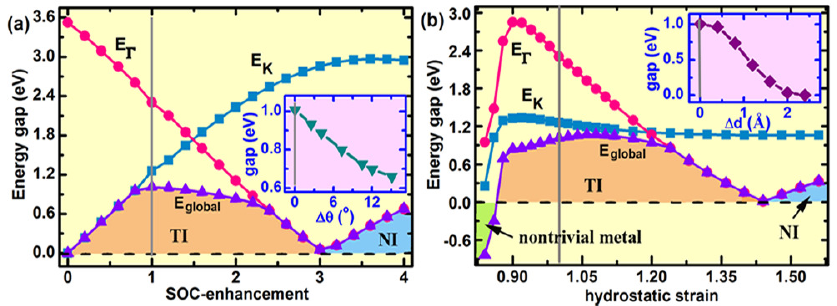}
\caption{(color online) The calculated energy gaps at $\Gamma$ point ($E_\Gamma$), $K$ point ($E_K$), and the global energy gap ($E_{\rm global}$) of H-Bi(111) as a function of SOC strength (a) and hydrostatic strain (b). A phase 
 transition from TI to normal insulator (NI) occurs accompanied by a gap closing and reopening at $\Gamma$. 
 Insets in panels (a) and (b) show  $E_{\rm global}$ versus $\theta_{\rm H-Bi-Bi}$ and $d_{\rm Bi-H}$, respectively.} 
\label{bandgap}
\end{figure}

To utilize the QSH effect in room temperature QSH-based electronic devices, a large band gap is needed. As shown in 
Figs.~\ref{band}(b) and \ref{band}(d), an indirect band gap of 1.01~eV and a direct band gap of 0.34~eV are obtained for 
H-Bi(111) and H-Bi(\={1}10), respectively, which are large enough for practical applications at room temperature. Especially 
for H-Bi(111), the band gap is by far larger than those of known 2D and 3D TIs. In order to test the stability 
of such a giant band gap, we expose the electronic structure of the bilayers to variation of different parameters. 
Figure~\ref{bandgap}(a) shows the variation of the energy gap at the $\Gamma$ point ($E_\Gamma$), $K$ point ($E_K$), and the global energy gap ($E_{\rm global}$) of H-Bi(111) as the SOC strength, $\lambda$, is varied. Such a variation can be realized experimentally by alloying Bi with isoelectronic Sb~\cite{Hsieh}. Starting from a calculation without SOC, it can be seen that $E_\Gamma$ closes with increasing $\lambda$, while $E_K$ opens accordingly. The transition from a direct ($E_K = E_{\rm global}$) to an indirect band gap occurs when the relative SOC strength ($\lambda/\lambda_{0}$, where $\lambda_{0}$ is the actual SOC strength) exceeds $0.8$. At $\lambda_{0}$, the global energy gap reaches its maximum and stays rather constant, since both the highest occupied band at the $K$ point and the lowest unoccupied band at the $\Gamma$ point are down shifted with increasing $\lambda$. With further enhancing SOC, the band gap becomes direct but at the $\Gamma$ point, and  then decreases rapidly. A band gap closing and reopening occurs at $\lambda/\lambda_0 = 3.0$, marking thus a band inversion and phase transition from TI to normal insulator (NI) that is confirmed by our topological analysis.

The lattice constant $a$, the angle $\theta_{\rm H-Bi-Bi}$ (or bond-length $d_{\rm Bi-Bi}$), and the Bi-H separation 
$d_{\rm Bi-H}$ are three key structural parameters to define the lattice of H-Bi(111). They might be altered, e.g.\
by epitaxial constraints imposed by a substrate. To reveal their influence, we show the variation of the band gap as a 
function of strain $a/a_0$ where $a_0$ is the equilibrium lattice constant, $\Delta\theta$ (change of $\theta_{\rm H-Bi-Bi}$
from the equilibrium value), and $\Delta d$ (change of $d_{\rm Bi-H}$) in Fig.~\ref{bandgap}(b), insets of \ref{bandgap}(a), 
and insets of \ref{bandgap}(b), respectively.  The large ($ > 0.6$~eV) indirect band gap is robust for $a/a_0$ ranging 
from 0.86 to 1.24, showing high adaptability in various application environments. When further compressed or expanded, 
the global band gap decreases rapidly. Under compression we observe the transition to a non-trivial metal, while with 
expansion a phase transition from TI to NI occurs when the gap closes and reopens at the $\Gamma$ point at $a/a_0 = 1.44$,
similarly to the case of strong SOC-enhancement. With applying hydrostatic strain, both $\Delta\theta$ and $\Delta$d are determined from a full optimization of the internal atomic coordinates. As shown in the insets of Fig.~\ref{bandgap}, the global band gap decreases with increasing of both $\Delta\theta$ and $\Delta d$. 

Having established the existence of a stable QSH phase, we focus now on the possibility to realize the QAH effect in 
functionalized Bi films. The essential ingredient for the transition from the QSH to the QAH phase is the breaking of 
TRS, since the QSH phase can be considered as two copies of QAH states that are coupled together by 
TRS~\cite{yu}. To date, most of the research on magnetically doped TIs has been focused on 
transition-metal doping~\cite{liu,yu,chang,niu}. In the following, we show theoretically that the realization of 
ferromagnetic ordering and QAH states without transition-metal doping can be achieved when the hydrogen atoms are removed 
from one side of H-Bi(111), while keeping the other side hydrogenated (what we call half-hydrogenated Bi or semihydrogenated Bi). 
Experimentally, semihydrogenated layers might be accessible when one side of the Bi film is protected by a substrate
and the other one is exposed to hydrogen. For graphene, half-hydrogenated layers have been successfully 
fabricated~\cite{Haberer}.

\begin{figure}
\centering
\includegraphics{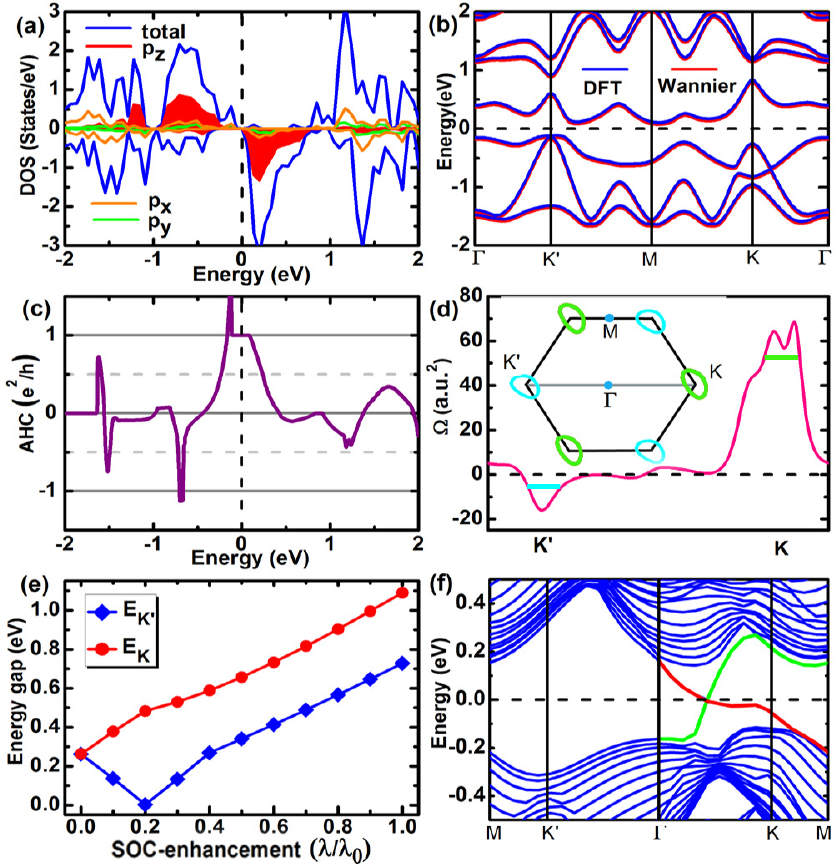}
\caption{(color online)
 (a) Total density of states (DOS) and partial DOS of the unhydrogenated Bi atoms (decomposed into 6p$_x$, 6p$_y$, 
 and 6p$_z$ states) of half H-decorated Bi(111) without SOC. Positive and negative values indicate spin-up and 
 spin-down channels, respectively. (b) Wannier and first-principles band structures with SOC for half H-decorated Bi(111). 
 The Wannier results have been shifted down by 30~meV for visibility.
 (c) Anomalous Hall conductivity as a function of the position of the Fermi level $E_{\rm F}$. (d) Berry curvature distribution
 of the occupied bands in $K-\Gamma-K^\prime$ direction. Inset shows the contour of Berry curvature distribution (as marked in main panel) around valleys $K$ and $K^\prime$. (e) The energy gaps at valleys $K (E_K)$ and $K^\prime (E_{K^\prime})$ as a function of SOC strength. (f) Band structures of zigzag-terminated half H-decorated Bi(111) exhibiting the valley-polarized QAH states. The states located at different edges are indicated by different colours.}
\label{mag}
\end{figure}

To understand the origin of the spin-polarization which we encounter, the calculated total density of states (DOS) and partial 
DOS for semihydrogenated Bi(111) without SOC are shown in Fig.~\ref{mag}(a). We clearly see that the spin-polarization
is mainly carried by the p$_z$ states of the unhydrogenated Bi atoms due to the breaking of the orbital network (see also 
spin density plots in supplementary material~\cite{sup}). The Fermi level is pinned inside the gap between the filled spin-up and 
empty spin-down Bi 6p$_z$ states, leading to a magnetic moment of $1.0 \mu_B$ per unit cell. Furthermore, non-spinpolarized DFT 
calculations for semihydrogenated Bi(111) show that the spin-polarized state is more stable by about 93.3~meV, proving that
the ground state  is magnetic. The calculated band structure in the presence of SOC is shown in 
Fig.~\ref{mag}(b), from which we conclude that our system is a magnetic insulator with the sizable indirect energy gap of 
0.35~eV. Calculation of the magnetic anisotropy energy shows that the easy magnetization axis points out of plane and it is by 1.19~meV lower in energy than the in-plane spin orientation. The existence of an insulating, magnetic ground state does not necessarily result in a
strong magnetic coupling and furthermore in a QAH phase. To access the nature of exchange coupling between the 
spin moments of unhydrogenated Bi atoms, which are separated by  5.49{\AA}, we compute the total energies of ferromagnetic ($E_{\rm FM}$) and antiferromagnetic ($E_{\rm AFM}$) states, finding  that  the ferromagnetic order is favored by $\Delta E = E_{\rm AFM} - E_{\rm FM}$ of 21.79~meV~\cite{sup}.

To identify the topological properties and predict a stable QAH state resulting from  the sizable energy gap, we calculate 
the anomalous Hall conductivity $\sigma_{xy}=(e^{2}/h)\mathcal C$, where $\mathcal C$, quantized and known as the first Chern
number in case of an insulator, can 
be obtained as an integral of the  Berry curvature of occupied state $\Omega({\bf k})$ over the Brillouin zone~\cite{sup}. The anomalous Hall conductivity as a function of band filling is calculated and presented 
versus the position of the Fermi level in Fig.~\ref{mag}(c). When the chemical potential is located within the energy gap, the Chern number of all occupied states indeed acquires an integer value of $+1$, confirming the QAH effect in semihydrogenated Bi(111). However, the semihydrogenation of Bi(111) leads to the breaking of TRS and inversion symmetry simultaneously, and band structures at valleys $K$ and $K^\prime$ have different patterns (Fig.~\ref{mag}(b)). Valleys $K$ and $K^\prime$ are distinguishable and the valley-polarized QAH state, which exhibits properties of both QAH state and quantum valley Hall (QVH) state~\cite{dxiao,pan}, is obtained. For further insight, the Berry curvature of all occupied bands along $K-\Gamma-K^\prime$ path is plotted in Fig.~\ref{mag}(d). It is clearly visible that the Berry curvature distribution is localized in the vicinity of $K$ and $K^\prime$ and it has opposite sign around the two valleys. The evolution of the energy gap at $K$ and $K^\prime$, shown in Fig.~\ref{mag}(e) as a function of the SOC strength, reveals the underlying physics of formation of the QVH state. At $K^\prime$, the energy gap closes and reopens as the SOC is increased, while that at the $K$-point always opens, indicating that a topological phase transition occurs at the $K^\prime$ but not at $K$, resulting in different valley-resolved Chern numbers, i.e., $\mathcal C_{K} = 1$ and $\mathcal C_{K^\prime} = 0$. To further confirm the valley-polarized QAH state, edge states of zigzag-terminated half H-decorated Bi(111) at valleys $K$ and $K^\prime$ are calculated and are shown in Fig.~\ref{mag}(f). The number of edge states in which valley indeed corresponds to the corresponding valley Chern number.

For device applications, it is important to make sure that, given a large enough bulk 
band gap of the lattice-matching substrate, which is aligned with the Bi-originated 2D gap, the predicted topological properties are
preserved~\cite{zhoum}. While owing to the enlarged lattice constant both Bi(111)~\cite{Drozdov} and Bi chalcogenides~\cite{Hirahara,zfwang,Kim} are not suitable for the purpose, we demonstrate this taking MoS$_2$ ($\sqrt{3}\times\sqrt{3}$) as an example substrate, 
which fits hydrogenated Bi nicely both in lattice constant as well as in alignment of the band gaps. We confirmed that magnetism for the semi-hydrogenated Bi is maintained and the topological properties for both full- and semi-hydrogenated cases are unchanged~\cite{sup}.

Non-trivial topological QAH states occur also for not fully hydrogenated Bi(\={1}10) under strain~\cite{sup}. Removing half 
of the hydrogen atoms in one of two layers results in a magnetic ground state, but this state shows metallic character with 
a global energy gap of $-0.46$~eV at the equilibrium lattice constant. Under hydrostatic strain, a transition from metal to 
insulator occurs, leading to the realization of the QAH state. Similar to the H-Bi(111), the QSH states form in the F-, 
Cl-, Br-, and I-decorated Bi(111), with gigantic energy gaps of 1.10~eV, 0.93~eV, 0.88~eV, and 0.87~eV, respectively~\cite{sup}. For half decorated cases, a small insulating gap of 11.57~meV exists and a QAH phase with the Chern number ${\cal C}=+1$ is obtained in half I-decorated Bi(111), while the half F-, Cl-, and Br-decorated films have metallic character. Like in half-hydrogenated Bi(\={1}10), the band gaps are sensitive to hydrostatic strain and the QAH states can be induced in most cases by moderate strain~\cite{sup}.

In summary, by performing DFT calculations for fully and semi-hydrogenated Bi(111) and Bi(\={1}10), we have demonstrated that both QSH and QAH states with giant band gaps can be realized. Especially in H-decorated Bi(111)
the band gaps for the QSH and QAH states reach 1.01 and 0.35~eV, respectively, which is larger than any known up to date system with corresponding topological properties. We further predict that  the QAH state in semi-hydrogenated Bi(111) is quite different from the normal one. It exhibits the properties of both QAH state and QVH state, realising a new quantum state, called valley-polarized QAH insulator. Both the QSH and valley-polarized QAH states survive even if the decorated Bi bilayers are on an appropriate substrate. Our results are of importance for further theoretical and experimental studies of topological insulators both from the point of view of fundamental exploration and as well as practical applications at room temperature.

\acknowledgments{We would like to thank Binghai Yan, Patrick Buhl, Jan-Philipp Hanke, Guillaume Geranton, and Frank Freimuth for useful discussions. This work was supported by the Priority Program 1666 of the German Research Foundation (DFG) and project VH-NG-513 of
the HGF. We acknowledge computing time on the supercomputers JUQUEEN and JUROPA at J\"{u}lich Supercomputing Centre and JARA-HPC of RWTH Aachen University.}

\end{document}